\documentclass[11pt, twoside]{article}
\pdfoutput=1
\usepackage{a4wide,amssymb,cite,graphicx}
\usepackage{epsfig}
\usepackage[usenames,dvipsnames]{color}
\usepackage{slashed}

\newcommand{\be}{\begin{equation}}
\newcommand{\ee}{\end{equation}}
\newcommand{\bea}{\begin{eqnarray}}
\newcommand{\eea}{\end{eqnarray}}

\begin{document}

\color{black}

\begin{flushright}
CTPU-15-26 \\
IPMU15-0214
\end{flushright}

\vspace{0.5cm}

\begin{center}
{\Large\bf\color{black} The diphoton resonance as a gravity mediator of dark matter}\\
\bigskip\color{black}\vspace{1.0cm}{
{\bf Chengcheng Han$^{1,*}$, Hyun Min Lee$^{2,\dagger}$, Myeonghun Park$^{3,\ddagger}$ and Ver\'onica Sanz$^{4,\sharp}$ }
\vspace{0.5cm}
} \\[8mm]

{\it $^1$ Kavli Institute for the Physics and Mathematics of the Universe (WPI), \\
Todai Institutes for Advanced Study, The University of Tokyo, Japan.} \\
{\it $^2$ Department of Physics, Chung-Ang University, Seoul 06974, Korea.}\\
{\it $^3$ Center for Theoretical Physics and Universe, Institute for Basic Science (IBS), \\
Daejeon 34051, Korea.}\\
{\it $^4$Department of Physics and Astronomy, University of Sussex, Brighton BN1 9QH, UK.  } \\
\end{center}
\bigskip
\centerline{\bf Abstract}
\begin{quote}
We consider the possibility of interpreting the recently reported diphoton excess at $750\,{\rm GeV} $ as a  spin-two massive particle (such as a Kaluza-Klein graviton in warped extra-dimensions) which serves as a mediator to Dark Matter via its gravitational couplings to the dark sector and to the Standard Model (SM).  We model non-universal couplings of the resonance to gauge bosons in the SM and to Dark Matter as a function on their localization in the extra dimension. We find that scalar, fermion or vector dark matter can saturate the dark matter relic density by the annihilation of dark matter into a pair of the SM particles or heavy resonances, in agreement with the diphoton resonance signal strength.  We check the compatibility of our hypothesis with other searches for the KK graviton. We show that the invisible decay rate of the resonance into a pair of dark matter is subdominant in the region of the correct relic density, hence leading to no constraints from the mono-jet bound at $8\,{\rm TeV}$ via the gluon coupling.  
We also discuss the kinematic features of the decay products of a KK graviton to distinguish the KK graviton from the SM backgrounds or a scalar particle interpretation of the diphoton resonance. 
\end{quote}

\vspace{2cm}

\begin{flushleft}
$^*$Email: chengcheng.han@ipmu.jp  \\
$^\dagger$Email: hminlee@cau.ac.kr \\
$^\ddagger$Email: parc.ctpu@gmail.com  \\
$^\sharp$Email: v.sanz@sussex.ac.uk
\end{flushleft}

\thispagestyle{empty}

\normalsize

\newpage

\setcounter{page}{1}

\section{Introduction}

Recently, both ATLAS data with $3.2\,{\rm fb}^{-1}$ \cite{atlas} and CMS data with $2.6\,{\rm fb}^{-1}$ \cite{cms} have shown a tantalizing hint for a diphoton resonance around $750\,{\rm GeV}$. The required production cross section for the diphoton resonance is about $6.2\pm 0.1\,{\rm fb}$ when Run 1 CMS and Run 2 CMS-ATLAS data are combined~\cite{veronica}. The compatibility of the signal hypothesis depends on the width of the resonance. Whereas CMS data prefers a narrow resonance, in the case of ATLAS, the local significance varies slightly from 3.6$\sigma$ for a narrow width, to 3.9$\sigma$ for a broader resonance of $45\,{\rm GeV}$. CMS has done a combination of their Run 1 and Run 2 data at around the resonance mass of $750\,{\rm GeV}$, finding  that their significance amounts to right above $3\sigma$. Accounting for the results from both experiments, a narrow resonance seems to be a more plausible hypothesis. Moreover, limits from null results in other searches such as $WW, ZZ, l{\bar l}$, $h h$ and dijets provide the information on other possible couplings of the resonance to SM particles. 

In order to prepare for more data coming in 2016, it is useful to explore scenarios for explaining the nature of the resonance. There have already appeared a large amount of recent papers explaining the diphoton resonance in various contexts beyond the SM~\cite{others}. In this article, we propose a different explanation, namely, that the resonance is a spin-2 mediator of Dark Matter. The spin-two particle will be referred as the Kaluza-Klein {\it KK-graviton}, although a glueball bound state of new strong interactions would lead to the same phenomenology~\cite{dual}.  

We revisit the Gravity-Mediated Dark Matter scenario suggested by some of the authors in the past where the massive KK-graviton in the warped extra dimension is responsible for mediating between dark matter and the SM particles \cite{gmdm}. We identify the KK graviton as the diphoton resonance observed at $750\,{\rm GeV}$ and introduce the interactions of the KK gravitons to transverse modes of SM gauge bosons and dark matter in a form of energy-momentum tensor.
Depending on the localization of bulk fields, the KK graviton interactions can be non-universal.
We assume that SM gauge bosons propagate in the bulk while dark matter is localized on the IR brane in the RS model.  The rest of the SM particles are assumed to be localized on the UV brane.
In this framework, the KK graviton has a naturally large coupling to dark matter, so it could decay dominantly into a pair of dark matter.  In this case, the invisible decay of the KK graviton can be constrained by mono-jet $+$ MET searches at the LHC.
Depending on whether dark matter is scalar, fermion or vector field, we consider the bound from the relic density of dark matter and show how much the invisible decay rate of the KK graviton is achieved,

The paper is organized as follows.
We begin with the description of the interactions of the KK graviton to the SM gauge bosons and dark matter.
Then, we summarize the status of the diphoton excesses at $750\,{\rm GeV}$ and bounds from other searches and constrain the couplings of a KK graviton to the SM gauge bosons.  Next, introducing the couplings of dark matter of arbitrary spin to the KK graviton, we search the parameter space of dark matter mass and coupling that are consistent with the  correct relic density. We also briefly discuss the direct and indirect detections of dark matter in the model and suggest using the angular distribution of photons to discriminate the spin of the diphoton resonance. 
Finally, conclusions are drawn.

\section{Interactions of the KK graviton }

The KK graviton $G_{\mu\nu}$ with mass $m_G$ has interactions to the SM gauge bosons and dark matter as
 \bea
{\cal L}_{\rm KK}
&=&-\frac{1}{\Lambda}G^{\mu\nu}\bigg[T^{\rm DM}_{\mu\nu}
+\sum_{a=1}^3c_a\left(\frac{1}{4}g_{\mu\nu} F^{\lambda\rho}_a F_{\lambda\rho,a}-F_{\mu\lambda,a}F^\lambda\,_{\nu,a}\right) \bigg]  \label{lag}
\eea
where the energy-momentum tensor for dark matter, depending on the spin ($0,1/2,1$) of dark matter, is given by
\bea
T^{\rm S}_{\mu\nu}&=&c_S\bigg[ \partial_\mu S \partial_\nu S-\frac{1}{2}g_{\mu\nu}\partial^\rho S \partial_\rho S+\frac{1}{2}g_{\mu\nu}  m^2_S S^2\bigg],\\
T^{\rm F}_{\mu\nu}&=& c_\chi \bigg[\frac{i}{4}{\bar\chi}(\gamma_\mu\partial_\nu+\gamma_\nu\partial_\mu)\chi-\frac{i}{4} (\partial_\mu{\bar\chi}\gamma_\nu+\partial_\nu{\bar\chi}\gamma_\nu)\chi-g_{\mu\nu}(i {\bar\chi}\gamma^\mu\partial_\mu\chi- m_\chi {\bar\chi}\chi) \bigg]
\nonumber \\
&&+\frac{i}{2}g_{\mu\nu}\partial^\rho({\bar\chi}\gamma_\rho\chi)\bigg],  \nonumber \\
T^{\rm V}_{\mu\nu}&=&
c_X\bigg[ \frac{1}{4}g_{\mu\nu} X^{\lambda\rho} X_{\lambda\rho}-X_{\mu\lambda}X^\lambda\,_{\nu}+m^2_X\Big(X_{\mu} X_{\nu}-\frac{1}{2}g_{\mu\nu} X^\lambda  X_{\lambda}\Big)\bigg].
\eea

Since the KK graviton is localized toward the IR brane, the KK graviton has unsuppressed couplings to the fields localized on the IR brane or in the bulk, but  suppressed couplings to the fields localized on the UV brane.  Thus, we assume that the SM gauge bosons are in the bulk and dark matter is localized on the IR brane whereas the rest of the SM particles including the Higgs doublet are localized on the UV brane. 
Then, the zero modes of transverse components of bulk gauge bosons couple to the KK graviton with sizable strength while the longitudinal components of bulk gauge bosons stemming from the Higgs doublet localized on the UV brane have suppressed couplings to the KK graviton. 
Couplings to the constant zero mode of bulk gauge fields in the Randall-Sundrum model \cite{rsmodel} are given by a volume suppression factor as $c_a\sim 1/(\ln(M_P/M_{IR}))$ with $M_{IR}$ being the IR brane scale. For instance, we get $c_a\sim 0.03$ for $M_{IR}\sim {\rm TeV}$ and $c_a\sim 0.1$ for $M_{IR}\sim 10^{14}\,{\rm GeV}$. On the other hand, when dark matter is localized towards the IR brane, the DM couplings to the KK graviton, $c_S, c_\chi, c_X$, become of order one. 
But, the precise values of DM couplings depend on the localization of bulk gauge fields and the warped factor in general warped geometries \cite{localization}, so we treat them to be independent parameters.

Having in mind the solution to the hierarchy problem by a warped factor, however, we can take the Higgs doublet and heavy quarks to be localized on the IR brane too.
In this case, the longitudinal components of bulk gauge bosons have sizable couplings to the KK graviton and there are more decay modes of the KK graviton such as $t{\bar t}, hh$, leading to more channels to test the scenario of the KK graviton.
But, for simplicity, we focus on the minimal model of the KK graviton with couplings only to transverse components of gauge bosons and dark matter of arbitrary spin, $s=0,1/2, 1$, for explaining the diphoton resonance at $750\,{\rm GeV}$.  On the other hand, as the Higgs doublet is localized on the UV brane, the Higgs portal couplings between dark matter and Higgs doublet, such as $|H|^2S^2$ for scalar dark matter, are suppressed. 

In the basis of gauge boson mass eigenstates in the SM, the KK graviton couplings to the gauge bosons become
\bea
{\cal L}^V_{\rm KK}&=&-\frac{1}{\Lambda}G^{\mu\nu} \bigg[ c_{\gamma\gamma}\left(\frac{1}{4}g_{\mu\nu} A^{\lambda\rho} A_{\lambda\rho}-A_{\mu\lambda}A^\lambda\,_{\nu}\right) + c_{Z\gamma} \left(\frac{1}{4}g_{\mu\nu} A^{\lambda\rho} Z_{\lambda\rho}-A_{\mu\lambda}Z^\lambda\,_{\nu}\right)  \nonumber \\
&&+c_{ZZ}\left(\frac{1}{4}g_{\mu\nu} Z^{\lambda\rho} Z_{\lambda\rho}-Z_{\mu\lambda}Z^\lambda\,_{\nu}\right)+c_{WW}\left(\frac{1}{4}g_{\mu\nu} W^{\lambda\rho} W_{\lambda\rho}-W_{\mu\lambda}W^\lambda\,_{\nu}\right)  \nonumber \\
&&+c_{gg}\left(\frac{1}{4}g_{\mu\nu} G^{\lambda\rho} G_{\lambda\rho}-G_{\mu\lambda}G^\lambda\,_{\nu}\right) \bigg]
\eea
where
\bea
c_{\gamma\gamma}&=& c_1\cos^2\theta_W + c_2 \sin^2\theta_W, \nonumber \\
c_{Z\gamma} &=& (c_2-c_1) \sin(2\theta_W),  \nonumber \\
c_{ZZ} &=& c_1 \sin^2\theta_W + c_2 \cos^2\theta_W,  \nonumber \\
c_{WW}&=& 2c_2,  \nonumber \\
c_{gg}&=& c_3. \label{c3}
\eea
When $c_1=c_2$, the $Z\gamma$ decay mode is absent. As it is expected that the spin-2 resonance is limited similarly to the spin-0 resonance by the $Z\gamma$ searches that give rise to $\sigma(gg\rightarrow S\rightarrow Z\gamma)<4\,{\rm fb}$ at LHC Run 1 \cite{zgamma8TeV}. Henceforth, we focus on the case with $c_1=c_2$.

Before closing the section, we remark on the couplings of the radion. The radion appears as a massless mode in the original RS model but it gets massive only after a stabilization mechanism is introduced. In the dual picture of strongly coupled dynamics in four dimensions, both spin-0 and spin-2 particles can appear as the lowest states, with similar couplings to the radion and KK-graviton, respectively \cite{gmdm}.  
The radion couples to the SM through the trace of the energy-momentum tensor, but the precise values of radion couplings depend on the overlap between the wave functions of the radion and the fields in the extra dimension,  similarly to the KK graviton. The important difference from the KK graviton is that the radion couplings to the transverse components of SM gauge bosons are  induced by trace anomalies so they are loop-suppressed. 
If the radion has a sizable coupling to gluons beyond the minimal setup, it might also be responsible for the $750$ GeV diphoton resonance.  When the radion is a mediator of dark matter, it could saturate the relic density through the annihilation of dark matter into a pair of the SM fermions or Higgs pair, leading to an interesting DM phenomenology \cite{gmdm}. The detailed discussion on the radion is beyond the scope of our work.    Therefore, we focus on the KK graviton in the following discussion.

\section{Diphoton resonance from the KK graviton}

We first discuss the requirements on the model from the observed diphoton resonance at $750\,{\rm GeV}$ in the LHC $13\,{\rm TeV}$. Then, we impose them on the KK graviton and show the parameter space that accommodates the diphoton resonance.

\subsection{Constraints from diphoton resonance}

The mass and width of the diphoton resonance is inferred from ATLAS data \cite{atlas} as
\bea
m_G&\approx&750\,{\rm GeV}, \\
\frac{\Gamma_G}{m_G}&\approx&0.06.
\eea
The production cross sections required to explain the reported  diphoton excesses in ATLAS ($3.2\,{\rm fb}^{-1}$) and CMS ($2.4\,{\rm fb}^{-1}$) at 13 TeV are, respectively,
\bea
\sigma(pp\rightarrow \gamma\gamma)_{\rm ATLAS}&=& (10\pm 3) \,{\rm fb}, \\
\sigma(pp\rightarrow \gamma\gamma)_{\rm CMS}&=& (6\pm 3) \,{\rm fb}. 
\eea
Henceforth, we are requiring the production cross section for the KK graviton to be $\sigma(pp\rightarrow \gamma\gamma)\approx 8\,{\rm fb}$ at $13\,{\rm TeV}$ from the averaged central values. 
We note that the gluon fusion process ($gg$) is better than the diquark process ($q{\bar q}$) for the resonance production,  due to a larger increase of the signal significance from Run 1 at 8 TeV to Run 2 at 13 TeV, and the gain factor is parametrized by the double ratio \cite{veronica}, ${\cal R}=(\sigma_S/\sqrt{\sigma_B})_{\rm 13TeV}/(\sigma_S/\sqrt{\sigma_B})_{\rm 8TeV}$ where $\sigma$ are the cross sections of signal($S$) and background($B$), so ${\cal R}_{gg}\simeq 3$ and ${\cal R}_{q{\bar q}}\simeq 1.7$, when the ratio of background cross sections is order 2.
The gain factor is largely insensitive to the spin and CP properties of the resonance \cite{veronica}.
It was also pointed out that the required cross section for the diphoton excess at $750\,{\rm GeV}$ is compatible with the excluded cross section translated from CMS Run 1, which is in the range of $2-8\,{\rm fb}$ \cite{veronica}. 
In Table \ref{tab:tab1}, we summarize the KK-graviton searches in different channels. Given enough signal events in the $\gamma\gamma$ channel, we also need to satisfy the limits from  8/13 TeV data.
For dijet bound in Table \ref{tab:tab1}, we note that the quoted limit is for $\sigma\times BR \times A$.

\begin{table}[h!]
\centering
\begin{tabular}{|c||c|c|c||c|}
\hline            ~~~~Channels~~~~ &  ~~~~$\sqrt{s}$=8 \rm{TeV}~~~~        &  ~~~~$\sqrt{s}$=13 \rm{TeV}~~~~         \\
\hline            $WW(lvjj)$      &    $\lesssim$ 68 fb \cite{Aad:2015ufa}          &	$\lesssim$ 259 fb \cite{WW}    \\
\hline            $ZZ(lljj)$      &    $\lesssim$  37 fb \cite{Aad:2014xka}          &   $\lesssim$ 151 fb \cite{ZZ}     \\
\hline            $\gamma\gamma$  & $\lesssim$  2.4 fb  \cite{Aad:2015mna}          &     $\lesssim$ 11 fb  \\
\hline             dijet          &    $\lesssim$  14 pb \cite{Aad:2014aqa}                   &       -          \\
\hline             monojet        &   $\lesssim$  270 fb \cite{Aad:2014nra}                      &       -           \\
\hline
\end{tabular}
\caption{Bounds from the KK graviton searches at the LHC.}
\label{tab:tab1}
\end{table}

The total cross section for the spin-two particle with couplings defined in Eq.~\ref{c3} can be computed for LHC13 energies
\bea
\sigma(pp\rightarrow G\rightarrow \gamma\gamma)= ( 8.6 \textrm{pb} )  \, \left( c_{gg}\frac{3 \textrm{TeV}}{\Lambda}\right)^2 \,  \left( c_{\gamma\gamma}\frac{3 \textrm{TeV}}{\Lambda}\right)^2 \, \left(\frac{\textrm{GeV}}{\Gamma_G} \right)   \label{crossection}
\eea
where $\Gamma_G$ is the total decay width of the KK graviton. 
In other words, we get
\be
\frac{c_{gg} \, c_{\gamma\gamma}}{(\Lambda/ \textrm{TeV} )^2}  \simeq \, 3 \times 10^{-3} \, (\Gamma_G/{\rm GeV})^{1/2} 
\ee
when we use $\sigma\sim 6$ fb. These numbers were obtained with a version of the RS model~\cite{RSfr} in {\tt Feynrules}~\cite{feynrules}, using {\tt UFO} format~\cite{UFO} and ran through {\tt Madgraph}~\cite{Madgraph}. 
In Table \ref{tab:tab2}, the production cross sections for the KK graviton at the LHC $8\,{\rm TeV}$ and $13\,{\rm TeV}$ are shown for a benchmark point of the gluon coupling. 
\begin{table}[t!]
\centering
\begin{tabular}{|c|c|c|c|c|}
\hline              ~~~~$\sqrt{s}$=8 \rm{TeV}~~~~        &  ~~~~$\sqrt{s}$=13 \rm{TeV}~~~~ &  ~~~~$\sigma_{13 \rm{TeV}}/\sigma_{8 \rm{TeV}}$~~~~    &  ~~~~$\Gamma_{G \rightarrow g g }$~~~~      \\
\hline                  105 fb       &	465 fb   & 4.4 & 0.015 \rm{GeV} \\
\hline
\end{tabular}
\caption{ Production cross section for $g g \rightarrow G$ at $\sqrt{s}$=8/13 \rm{TeV} LHC and the ratio of production cross sections. $\Lambda$=3 \rm{TeV} and $c_3=0.1$ are chosen for a benchmark point.}
\label{tab:tab2}
\end{table}

\subsection{KK graviton couplings }

From the interactions of the KK graviton given in eq.~(\ref{lag}), the total decay rate of the KK graviton $\Gamma_G$ is computed as the sum of decay rates for a pair of gauge bosons
\bea
\Gamma(\gamma\gamma)&=& \frac{c^2_{\gamma\gamma} m^3_G}{80\pi\Lambda^2}, \\
\Gamma(Z\gamma)&=&\frac{c^2_{Z\gamma} m^3_G}{160\pi \Lambda^2} \Big(1- \frac{m^2_Z}{m^2_G}\Big)^3 \Big(1+\frac{m^2_Z}{2m^2_G}+\frac{m^4_Z}{6 m^4_G} \Big),\\
\Gamma(Z Z)&=& \frac{c^2_{ZZ} m^3_G}{80\pi \Lambda^2}\Big(1- \frac{4m^2_Z}{m^2_G}\Big)^\frac{1}{2}
\Big(1-\frac{3m^2_Z}{m^2_G}+\frac{6m^4_Z}{m^4_G}\Big), \\
\Gamma(W W )&=& \frac{c^2_{WW} m^3_G}{160\pi \Lambda^2}\Big(1- \frac{4m^2_W}{m^2_G}\Big)^\frac{1}{2}
\Big(1- \frac{3m^2_W}{m^2_G}+\frac{6m^4_W}{m^4_G}\Big), \\
\Gamma(gg)&=&  \frac{c^2_{gg} m^3_G}{10\pi\Lambda^2}.
\eea
and the invisible decay rate into a pair of dark matter, depending on the spin of dark matter, given as follows,
\bea
\Gamma(SS)&=&  \frac{c^2_S m^3_G}{960 \pi \Lambda^2} \Big(1-\frac{4m^2_S}{m^2_G}\Big)^\frac{5}{2}, \\
\Gamma(\chi{\bar\chi})&=& \frac{ c_\chi^2 m^3_G}{160 \pi \Lambda^2}
 \left(1-\frac{4m^2_\chi}{m^2_G} \right)^\frac{3}{2} \left(1+\frac{8}{3} \frac{m^2_\chi}{m^2_G}\right), \\
 \Gamma(XX)&=& \frac{ c^2_X m^3_G}{960\pi \Lambda^2}\Big(1- \frac{4m^2_X}{m^2_G}\Big)^\frac{1}{2}
\Big(13+\frac{56m^2_X}{m^2_G}+\frac{48m^4_X}{m^4_G}\Big).
\eea

\begin{figure}
  \begin{center}
   \includegraphics[height=0.50\textwidth]{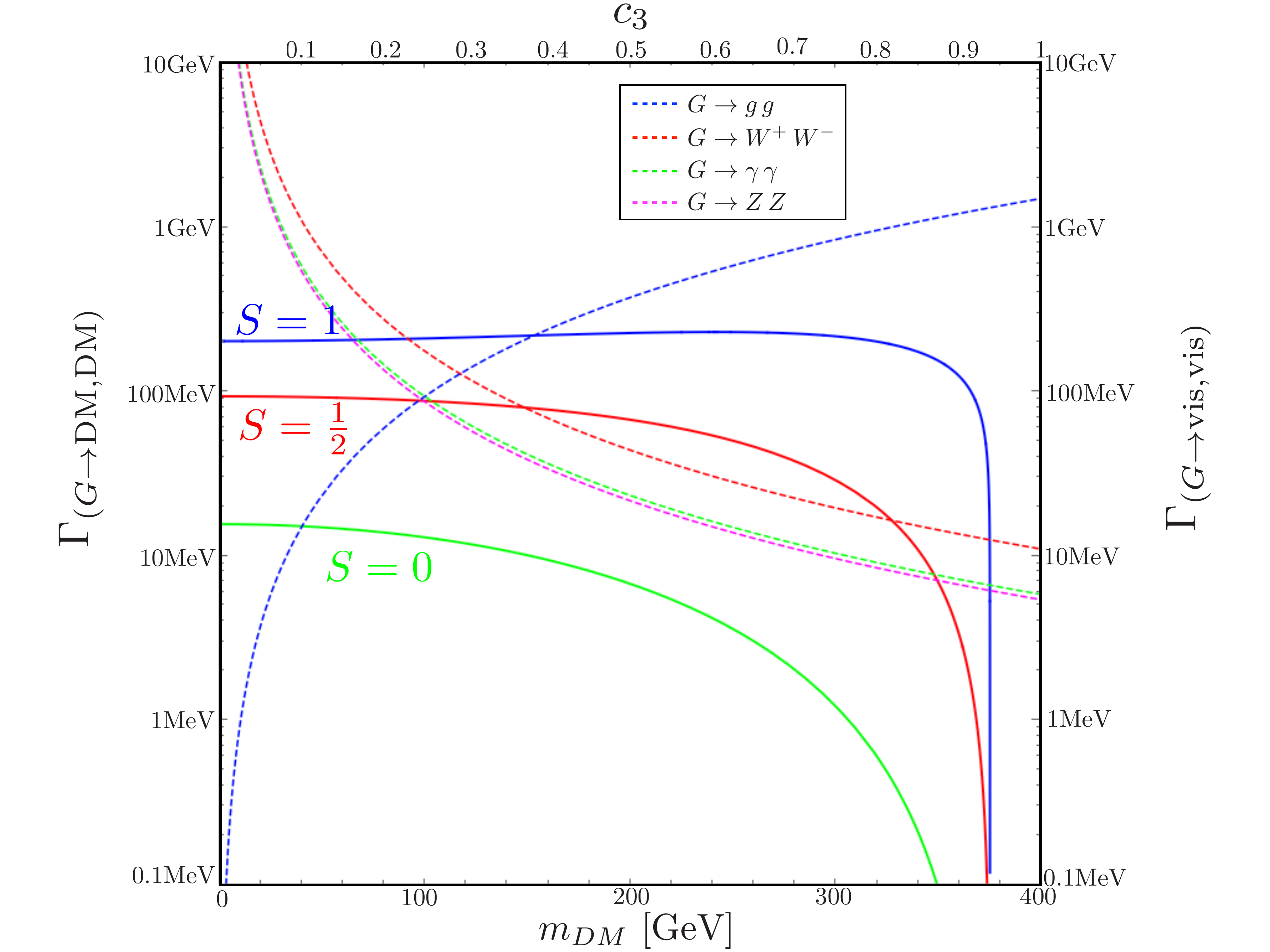}
   \end{center}
  \caption{Decay rates for the visible and invisible decays of the KK graviton as a function of $c_3$ and dark matter mass, respectively. We have set $\Lambda=3\,{\rm TeV}$, $m_G=750\,{\rm GeV}$ and the DM coupling is chosen as the same value, $c_{\rm DM}=1$. For visible decays, we imposed $c_1=c_2$ and the signal rate for $\Gamma_G=45\,{\rm GeV}$ for illustration. }
  \label{invisible}
\end{figure}

First we discuss the implications of the diphoton signal rate for the KK graviton couplings.
Taking the diphoton signal rate given in eq.~(\ref{crossection}) to be $\sigma\approx 6\,{\rm fb}$ and for $c_1=c_2$, the KK couplings are constrained to
\be
|c_1\cdot c_3| \approx 0.18 \Big(\frac{\Lambda}{3\,{\rm TeV}} \Big)^2\Big(\frac{\Gamma_G/m_G}{0.06}\Big)^{1/2}.
\ee

We have shown in Fig.~\ref{invisible} the invisible decay rate of the KK graviton as a function of DM mass in each case of DM spin.  As a result, we find that vector dark matter has the largest invisible decay rate for the same DM coupling. 
For $m_{\rm DM}\ll m_G=750\,{\rm GeV}$, the couplings of the KK graviton to dark matter  can be written as a function of the invisible decay rate approximately as follows,
\bea
 |c_S| &\approx& \Big(\frac{\Lambda}{3\,{\rm TeV}}\Big)\Big(\frac{\Gamma_{\rm inv}/m_G}{2.1\times 10^{-5}}\Big)^{1/2}
, \\
 |c_\chi| &\approx&   \Big(\frac{\Lambda}{3\,{\rm TeV}}\Big)\Big(\frac{\Gamma_{\rm inv}/m_G}{1.1\times 10^{-4}}\Big)^{1/2}
, \\
 |c_X|&\approx& \Big(\frac{\Lambda}{3\,{\rm TeV}}\Big)\Big(\frac{\Gamma_{\rm inv}/m_G}{2.7\times 10^{-4}}\Big)^{1/2}
.
\eea  
From the above expressions, we can get a rough idea on how large the partial invisible decay rate is, depending on the spin of dark matter.

\begin{figure}
  \begin{center}
   \includegraphics[height=0.30\textwidth]{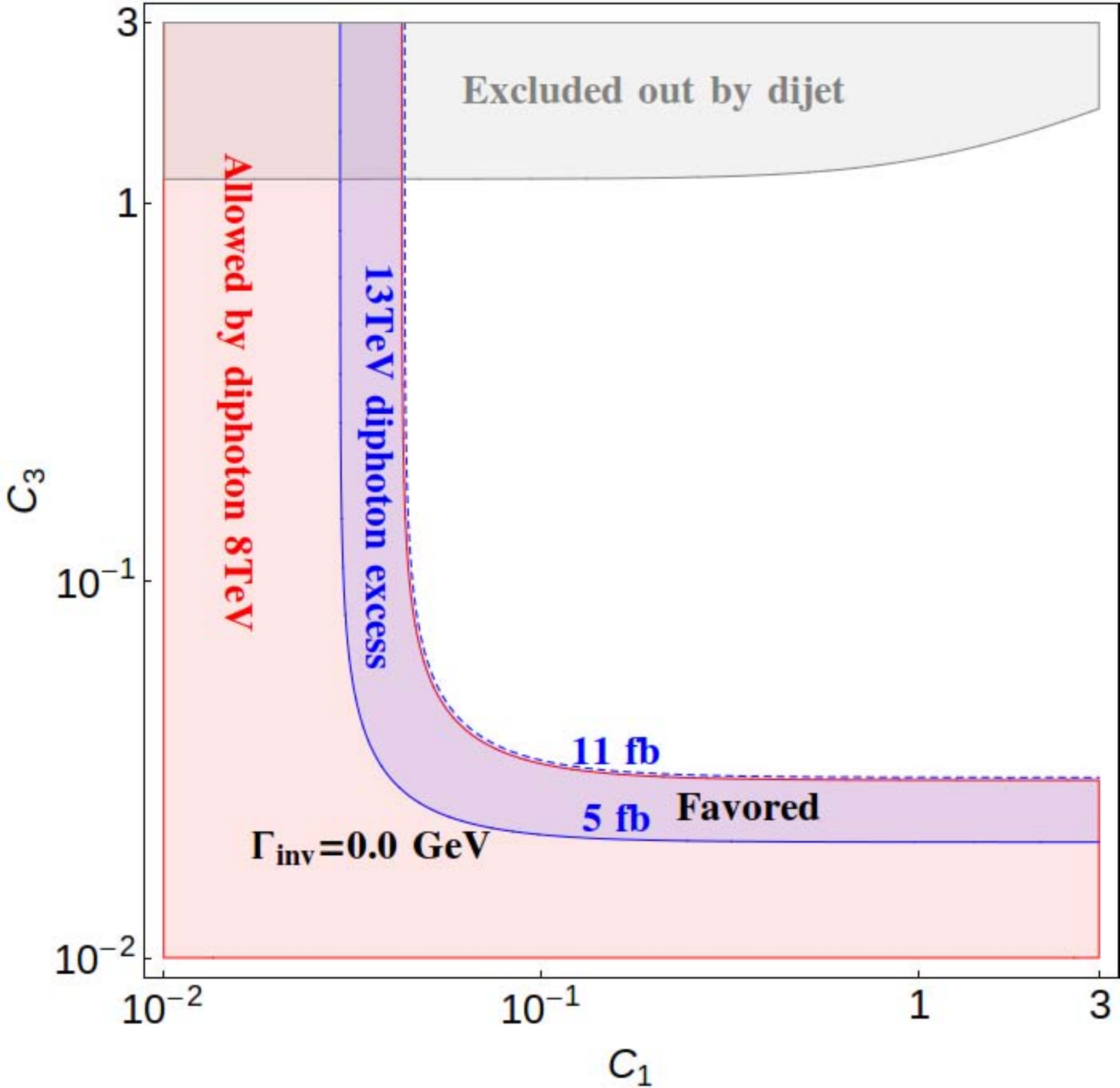}
    \includegraphics[height=0.30\textwidth]{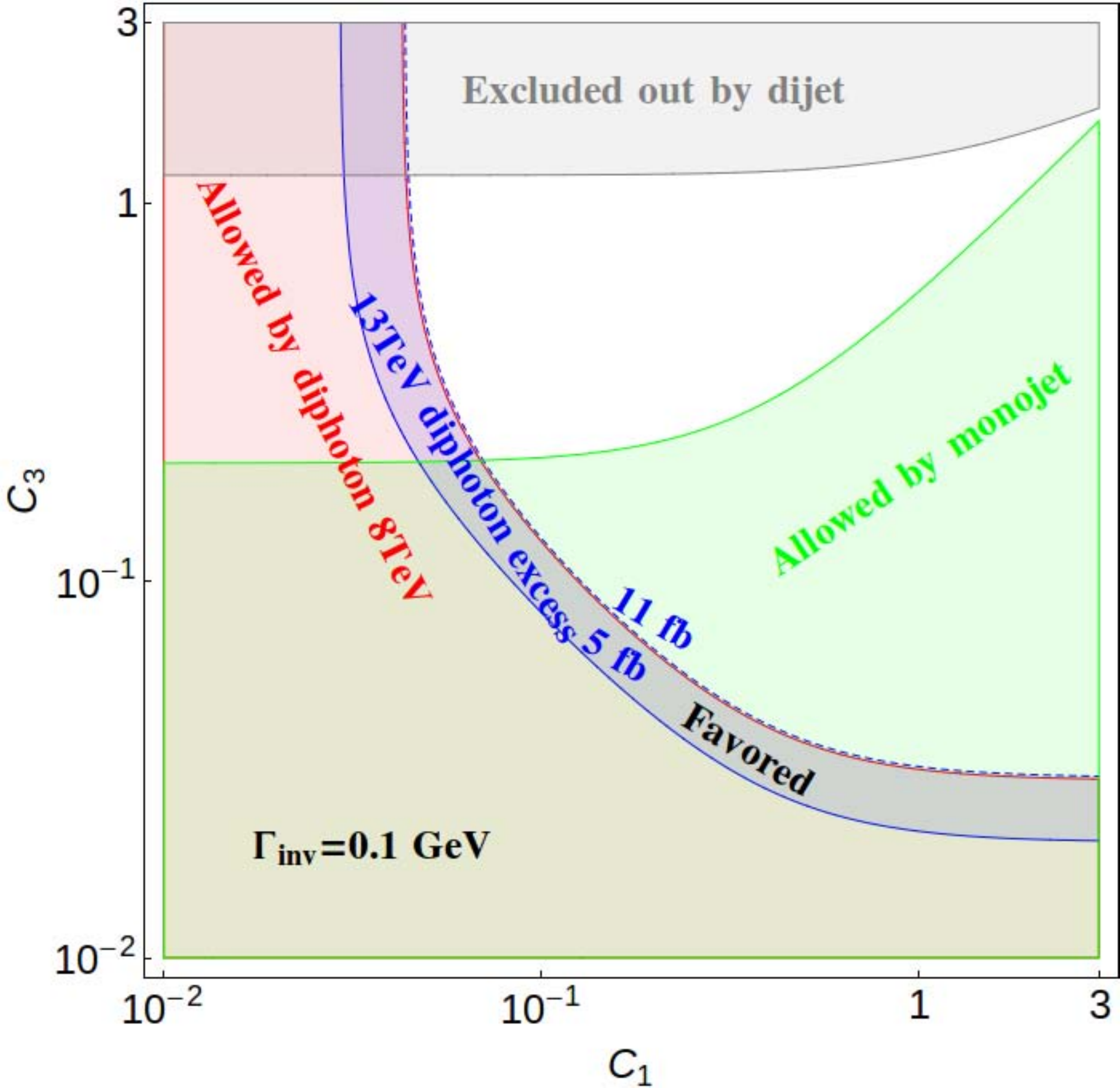} \includegraphics[height=0.30\textwidth]{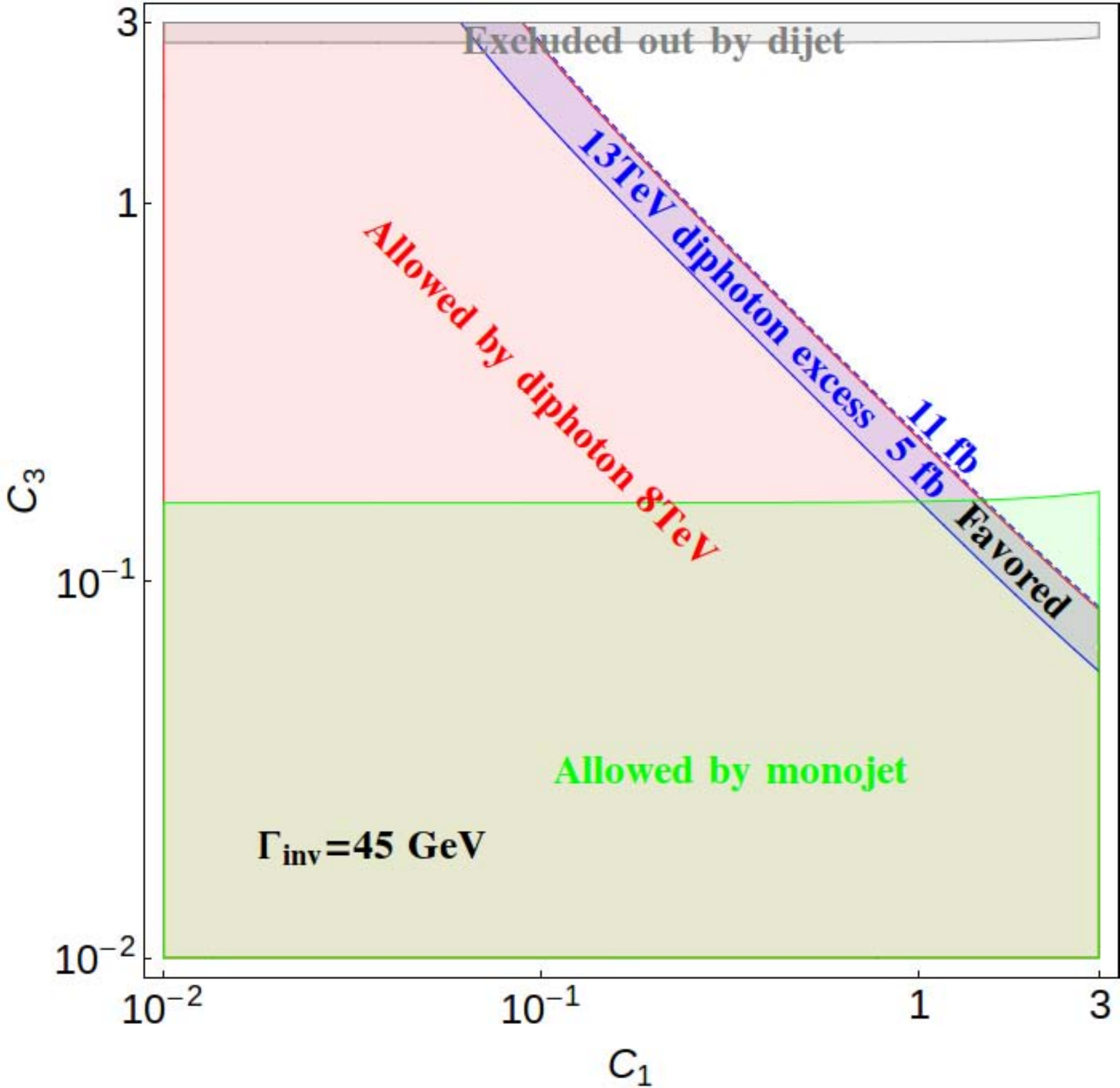}
   \end{center}
  \caption{Parameter space of $c_1$ and $c_3$ for the diphoton resonance. The region explaining the diphoton resonance at $750\,{\rm GeV}$ for $\sigma(pp\rightarrow\gamma\gamma)=5-11\,{\rm fb}$ is shown in blue.
  The diphoton limit from CMS $8\,{\rm TeV}$ is shown in red while the mono-jet limit from LHC $8\,{\rm TeV}$ is shown in green. The invisible decay width of the KK graviton is chosen to $\Gamma_{\rm inv}=0, 0.1, 45\,{\rm GeV}$, from left to right plots. We note that $|c_3|<1.5$ is favored for dijet searches in all the cases.}
  \label{couplings}
\end{figure}

In Fig.~\ref{couplings}, we have shown the parameter space for $c_1$ and $c_3$, that is consistent with the diphoton resonance at $750\,{\rm GeV}$, when $c_1=c_2$ is imposed. We imposed the production cross section for the diphoton resonance as $\sigma(pp\rightarrow\gamma\gamma)=5-11\,{\rm fb}$, depending on the invisible decay width of the KK graviton, $\Gamma_{\rm inv}=0, 0.1, 45\,{\rm GeV}$.  First, we found that the parameter space allowed by the production cross section for the diphoton resonance is consistent with the diphoton limit on the spin-2 resonance from CMS $8\,{\rm TeV}$ \cite{Aad:2015mna}. When the invisible width is comparable to the visible width, we also showed that the mono-jet $+$ MET limit  from LHC $8\,{\rm TeV}$ \cite{Aad:2014aqa} can constrain the parameter space further to smaller gluon couplings.
We note that if the total width of the KK graviton is $45\,{\rm GeV}$ as indicated in the best-fit of ATLAS data and it is dominated by the invisible decay mode, the mono-jet bound constrains the gauge couplings to $c_1\gtrsim 1.0$ and $c_3\lesssim 0.16$. This would mean that electroweak gauge bosons are localized towards the IR brane and gluons are delocalized from the IR brane. However, if the invisible decay rate is sub-dominant, the mono-jet searches do not constrain the gauge couplings, whereas dijet searches \cite{Aad:2014aqa} still constrain the gluon coupling to $|c_3|<1.5$. 
We should be open-minded to the choice of the total width until a further confirmation of the total width of the KK graviton is made. In particular, in the later discussion on dark matter, the invisible decay width turns out to be subdominant as compared to the visible decay width.

\section{Dark matter mediated by the KK graviton}

In this section, we consider the relic density condition in each case of dark matter and discuss the compatibility with the diphoton resonance and direct and indirect detection bounds. 

Dark matter can annihilate into a pair of the SM gauge bosons through the KK graviton in the s-channels and/or a pair of KK gravitons in the t-channel. 
When the thermal-averaged cross section for  dark matter is expanded in terms of the relative velocity as $\langle\sigma v\rangle=a+b v^2+c v^4$,  the relic density of dark matter is given by
\be
\Omega h^2=\frac{2.09\times 10^8\,{\rm GeV}^{-1}}{M_P \sqrt{g_{*s}(x_F)}(a/x_F+3b/x^2_F+20c/x^3_F)}
\ee
where $x_F=m_{DM}/T_F\simeq 20$ with  $T_F$ being the freeze-out temperature, $g_{*s}(x_F)$ is the effective number of relativistic species contributing to the entropy density at the freeze-out temperature.

\subsection{Scalar dark matter}

\begin{figure}
  \begin{center}
   \includegraphics[height=0.46\textwidth]{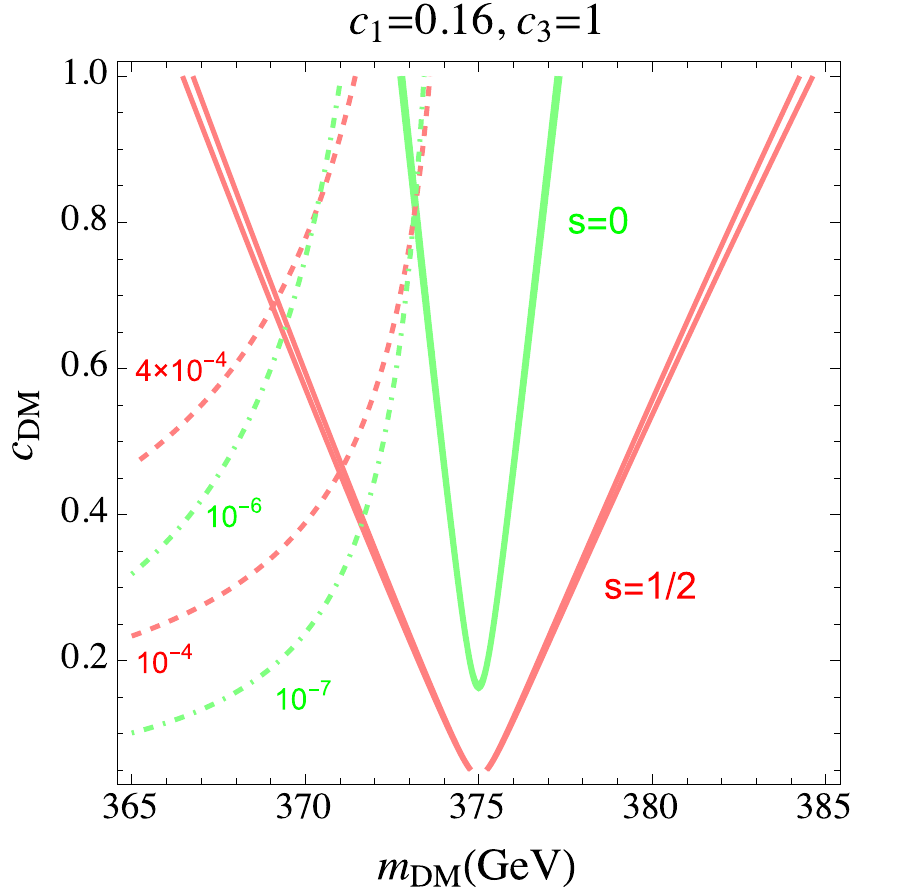}
    \includegraphics[height=0.46\textwidth]{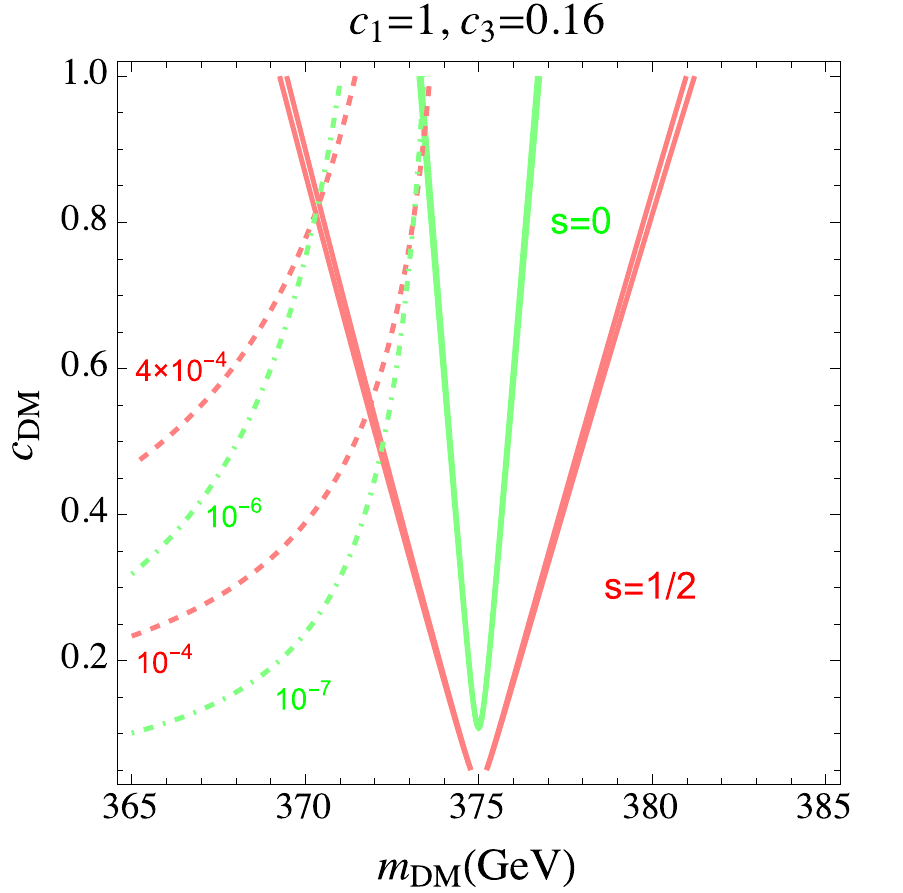}
   \end{center}
  \caption{Parameter space of $c_{\rm DM}$ and $m_{\rm DM}$ for spin-$0$ or $1/2$ dark matter, satisfying the relic density. Green and red solid lines satisfy the Planck $3\sigma$ band for the relic density.  Green dot dashed lines (red dashed lines) are contours with the same invisible decay width of the KK graviton in units of GeV. We have set $\Lambda=3\,{\rm TeV}$ and $m_G=750\,{\rm GeV}$. The KK couplings are chosen as $c_1=0.16, c_3=1$ on left and $c_1=1, c_3=0.16$ on right . }
  \label{scalarDM}
\end{figure}

The annihilation cross sections for scalar dark matter going into a pair of SM gauge bosons are given by
\bea
(\sigma v)_{SS\rightarrow Z Z}&\simeq& \frac{3c^2_S c^2_{ZZ}}{16\pi \Lambda^4} \frac{m^2_S m^4_Z}{(4m^2_S-m^2_G)^2+\Gamma^2_G m^2_G}\left(1-\frac{4m^2_S}{m^2_G}\right)^2\left(1-\frac{m^2_Z}{m^2_S}\right)^{\frac{1}{2}},  \\
(\sigma v)_{SS\rightarrow W W}&\simeq& \frac{3c^2_Sc^2_{WW}}{32\pi \Lambda^4} \frac{m^2_S m^4_W}{(4m^2_S-m^2_G)^2+\Gamma^2_G m^2_G}\left(1-\frac{4m^2_S}{m^2_G}\right)^2\left(1-\frac{m^2_Z}{m^2_S}\right)^{\frac{1}{2}}, \\
(\sigma v)_{SS\rightarrow \gamma\gamma}&\simeq&v^4 \cdot  \frac{c^2_S c^2_{\gamma\gamma} }{60\pi\Lambda^4}\frac{m^6_S}{(4m^2_S-m^2_G)^2+\Gamma^2_G m^2_G},\\
(\sigma v)_{SS\rightarrow gg}&\simeq& v^4 \cdot  \frac{2c^2_S c^2_{gg} }{15\pi\Lambda^4}\frac{m^6_S}{(4m^2_S-m^2_G)^2+\Gamma^2_G m^2_G}.
\eea
Thus, the annihilation of scalar dark matter into a pair of massless gauge bosons is d-wave suppressed. But, if the gluon coupling is the largest, the relic density is still determined dominantly by the annihilation into a pair of gluons. In Fig.~\ref{scalarDM}, we depict the parameter space of the DM coupling $c_{
\rm DM}=c_S$ and the DM mass $m_{\rm DM}=m_S$ by imposing the correct relic density from the Planck $3\sigma$ band \cite{planck}. For a large gluon coupling, the results show that there exists a wider region of the parameter space consistent with the relic density, near the resonance, $m_S=m_G/2$. We note that for the signal rate of the diphoton resonance with a narrow width, the  coupling between gauge bosons and the KK graviton should be taken to a smaller value. But, the result for the relic density does not change much as far as either $c_1$ or $c_3$ is of order one.  

When $m_S>m_G$, there is an extra contribution to the annihilation cross section, due to the t-channel for both models, as follows,
\bea
(\sigma v)_{SS\rightarrow GG} \simeq \frac{4 c_{S}^4 m_{S}^2}{9 \pi \Lambda^4 }
\frac{(1-r_S)^\frac{9}{2}}{r^4_S  (2-r_S)^2}   \label{tch-scalar}
\eea
with $r_S = \left(\frac{m_G}{m_S}\right)^2$.
Thus, the t-channel annihilation is s-wave, so it becomes dominant in determining the relic density for heavy scalar dark matter. But, for $m_G=750\,{\rm GeV}$, it turns out that there is no parameter space satisfying the relic density in this regime.

\subsection{Fermion dark matter}

The annihilation cross sections for fermion dark matter going into a pair of SM gauge bosons are also given by
\bea
(\sigma v)_{\chi{\bar\chi}\rightarrow ZZ} &\simeq&  v^2 \cdot \frac{c_\chi^2 c_{ZZ}^2}{144\pi \Lambda^4}
\frac{m_\chi^6}{(m_G^2-4 m_\chi^2)^2+\Gamma_G^2 m_G^2} \,\left(1-\frac{m_Z^2}{m_\chi^2}\right)^\frac{1}{2}\nonumber \\
&&\quad \times\Bigg(12-\frac{9m_Z^2}{m_\chi^2}+\frac{39m_Z^4}{8m_\chi^4}
-\frac{3m_Z^4}{m_G^2 m_\chi^2}+\frac{6m_Z^4}{m_G^4}
\Bigg)  , \\
(\sigma v)_{\chi{\bar\chi}\rightarrow WW} &\simeq&  v^2 \cdot \frac{c_\chi^2 c_{WW}^2}{288\pi \Lambda^4}
\frac{m_\chi^6}{(m_G^2-4 m_\chi^2)^2+\Gamma_G^2 m_G^2} \, \left(1-\frac{m_W^2}{m_\chi^2}\right)^\frac{1}{2}\nonumber \\
&&\quad\times\Bigg(12-\frac{9m_W^2}{m_\chi^2}+\frac{39m_W^4}{8m_\chi^4}
-\frac{3m_W^4}{m_G^2 m_\chi^2}+\frac{6m_W^4}{m_G^4}
\Bigg), \\
(\sigma v)_{\chi{\bar\chi}\rightarrow \gamma\gamma}&\simeq& v^2\cdot \frac{c^2_\chi c^2_{\gamma\gamma}}{12\pi\Lambda^4}\frac{m^6_\chi}{(4m^2_\chi-m^2_G)^2+\Gamma^2_G m^2_G},\\
(\sigma v)_{\chi{\bar\chi}\rightarrow gg}&\simeq&  v^2\cdot\frac{2c^2_\chi c^2_{gg}}{3\pi\Lambda^4}\frac{m^6_\chi}{(4m^2_\chi-m^2_G)^2+\Gamma^2_G m^2_G}.
\eea
We note that all the above annihilation channels are p-wave suppressed. As shown in Fig.~\ref{scalarDM}, the region of the parameter space of the DM coupling $c_{\rm DM}=c_\chi$ and the DM mass $m_{\rm DM}=m_\chi$ is wider than in the case of scalar dark matter, because the annihilation of fermion dark matter is less velocity-suppressed, although still being near the resonance with $m_\chi=m_G/2$.  

In Fig.~\ref{scalarDM}, we also overlay the contours of the same invisible decay rate of the KK graviton for both scalar and fermion dark matter in the same figure. In both cases, the invisible decay rate of the KK graviton is quite small in the region of the correct relic density, because of a large phase space suppression. We get that $\Gamma_{\rm inv}\sim 10^{-7}$ GeV for scalar dark matter and $\Gamma_{\rm inv}\sim 10^{-4}$ GeV for fermion dark matter.    
In this case, the decay rate of the KK graviton is determined mainly by the decay modes into SM gauge bosons.

When $m_{\chi}>m_G$, there is an extra contribution to the annihilation cross section, due to the t-channel for both models, as follows,
\bea
(\sigma v)_{\chi \bar\chi \rightarrow GG} &\simeq& \frac{c_{\chi}^4 m_{\chi}^2}{16 \pi \Lambda^4 }
\frac{(1-r_\chi)^\frac{7}{2}}{r^4_\chi (2-r_\chi)^2}  \label{tch-fermion}
\eea
with $r_\chi = \left(\frac{m_G}{m_\chi}\right)^2$. Then,   the t-channel annihilation is s-wave, so it becomes dominant in determining the relic density for heavy fermion dark matter.
But, similarly to the scalar dark matter case, for $m_G=750\,{\rm GeV}$, it turns out that there is no parameter space satisfying the relic density in this regime.

\subsection{Vector dark matter}

\begin{figure}
  \begin{center}
   \includegraphics[height=0.46\textwidth]{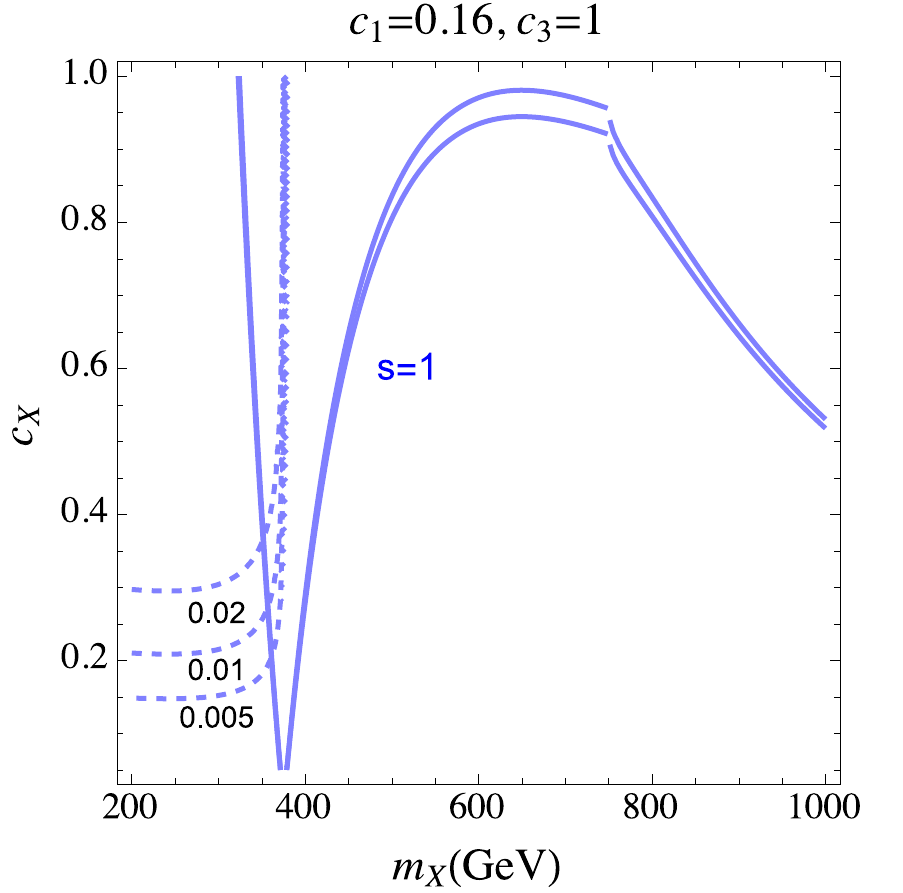}
    \includegraphics[height=0.46\textwidth]{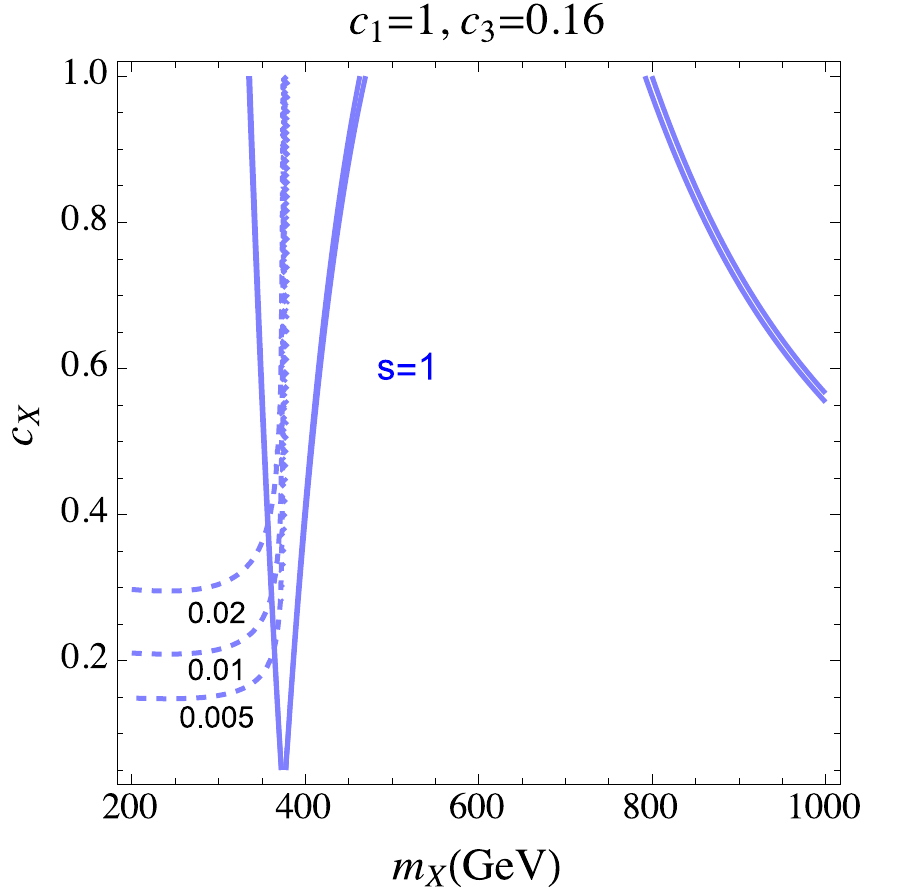}
   \end{center}
  \caption{Parameter space of $c_X$ and $m_X$ for vector dark matter, satisfying the relic density. Planck $3\sigma$ band is imposed for the relic density.  Blue solid lines satisfy the Planck $3\sigma$ band for the relic density.  Blue dashed lines are contours with the same invisible decay width of the KK graviton in units of GeV. We have set $\Lambda=3\,{\rm TeV}$ and $m_G=750\,{\rm GeV}$. The KK couplings are chosen as $c_1=0.16, c_3=1$ on left and $c_1=1, c_3=0.16$ on right. }
  \label{vectorDM}
\end{figure}

The annihilation cross sections for vector dark matter going into a pair of SM gauge bosons are also given by
\bea
(\sigma v)_{XX\rightarrow ZZ} &\simeq&  \frac{2 c_{X}^2 c_{ZZ}^2}{27\pi \Lambda^4}
\frac{m_{X}^6}{(m_G^2-4 m_{X}^2)^2+\Gamma_G^2 m_G^2}\, \left(1-\frac{m_Z^2}{m_{X}^2}\right)^\frac{1}{2}\nonumber \\
&&\quad\times\Bigg(12-\frac{9m_Z^2}{m_{X}^2}+\frac{147}{32}\frac {m_Z^4}{m_{X}^4}
-\frac{3}{4}\frac{m_Z^4}{m_G^2 m_{X}^2}+\frac{3}{2}\frac{m_Z^4}{m_G^4}
\Bigg)  , \\
(\sigma v)_{XX\rightarrow WW} &\simeq&  \frac{c_{X}^2 c_{WW}^2}{27\pi \Lambda^4}
\frac{m_{X}^6}{(m_G^2-4 m_{X}^2)^2+\Gamma_G^2 m_G^2}\, \left(1-\frac{m_W^2}{m_{X}^2}\right)^\frac{1}{2}  \nonumber \\
&&\times \Bigg(12-\frac{9m_W^2}{m_{X}^2}+\frac{147}{32}\frac {m_W^4}{m_{X}^4}
-\frac{3}{4}\frac{m_W^4}{m_G^2 m_{X}^2}+\frac{3}{2}\frac{m_W^4}{m_G^4}.
\Bigg), \\
(\sigma v)_{XX\rightarrow \gamma\gamma}&=&\frac{8c^2_X c^2_{\gamma\gamma}}{9\pi\Lambda^4}\frac{m^6_X}{(4m^2_X-m^2_G)^2+\Gamma^2_G m^2_G}, \\
(\sigma v)_{XX\rightarrow gg}&=&\frac{64c^2_X c^2_{gg}}{9\pi\Lambda^4}\frac{m^6_X}{(4m^2_X-m^2_G)^2+\Gamma^2_G m^2_G}.
\eea
In this case, all the above s-channels are s-wave, so they are relevant for both the early Universe at freeze-out temperature and the present Universe.  In Fig.~\ref{vectorDM}, it is shown that the correct relic density within the Planck $3\sigma$ band \cite{planck} can be obtained even away from the resonance with $m_X=m_G/2$, due to the s-wave nature of the annihilation processes. But, the mass of vector dark matter should be greater than about $300\,{\rm GeV}$ for the DM coupling $c_X$ less than unity.  Furthermore, as will be discussed further later, vector dark matter is most constrained by indirect detection experiments.

In Fig.~\ref{vectorDM},  we also overlay the contours of the same invisible decay rate of the KK graviton for vector dark matter in the same figure. The invisible decay rate of the KK graviton in the region of the correct relic density ranges between $\Gamma_{\rm inv}\sim 0.005-0.02$ GeV. In this case, the decay rate of the KK graviton can be dominated by the visible decay, so the mono-jet searches could not constrain the gluon coupling.

For $m_X>m_G$, there is an extra contribution to the annihilation cross section, due to the t-channel in both models, as follows,
\bea
(\sigma v)_{X X  \rightarrow GG} &\simeq&
\frac{c_{X}^4 m_{X}^2}{324 \pi \Lambda^4 }
\frac{\sqrt{1-r_X}}{r^4_X  (2-r_X)^2} \,
\bigg(176+192 r_X+1404 r^2_X-3108 r^3_X \nonumber \\
&&+1105 r^4_X+362 r^5_X+34 r^6_X \bigg)
\eea
with $r_X = \left(\frac{m_G}{m_X}\right)^2$.
Like the previous cases with other spins of dark matter, the t-channel annihilation of vector dark matter is s-wave, so it is as important as the s-channel annihilation for $m_X>m_G$.

\subsection{Direct and indirect detections of dark matter}

\begin{figure}
  \begin{center}
   \includegraphics[height=0.50\textwidth]{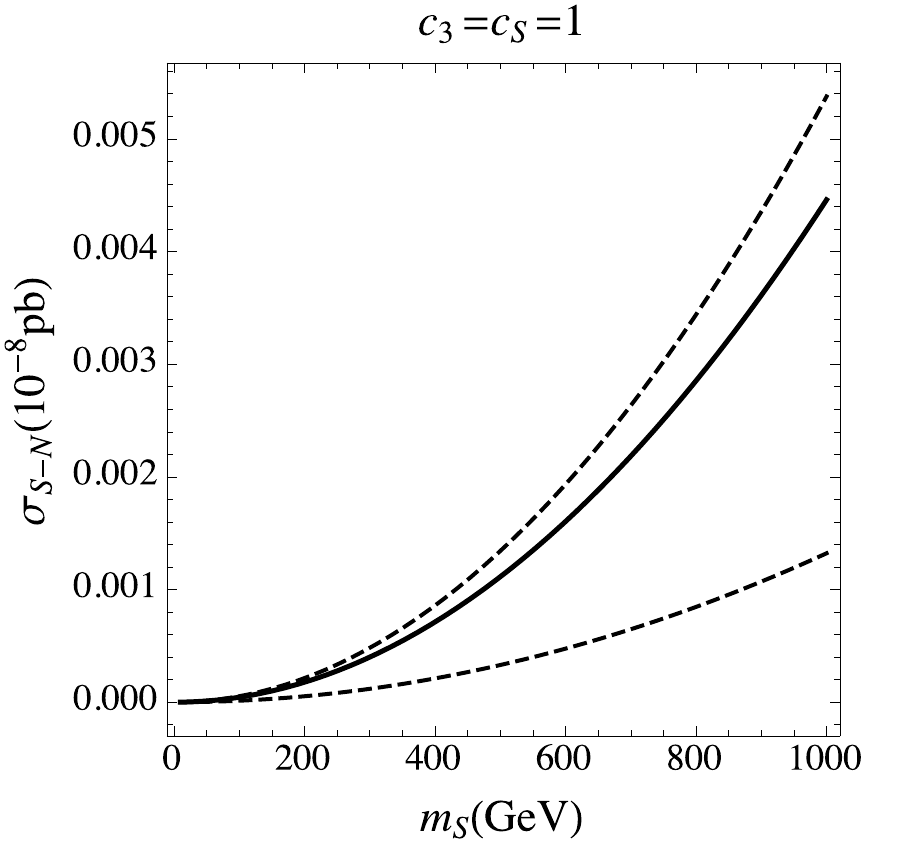}
       \end{center}
  \caption{DM-nucleon spin-independent scattering cross section as a function of DM mass for scalar dark matter, satisfying the relic density. We have set $\Lambda=3\,{\rm TeV}$ and $m_G=750\,{\rm GeV}$. Solid line is from the lattice result \cite{lattice} and dashed lines are given by the MILC results \cite{MILC}.}
  \label{direct}
\end{figure}

In our model, gluon interactions to dark matter can be sizable and dominant in determining the relic density for $m_{\rm DM}< m_G$. In this case, the corresponding gluon interactions are relevant for the direct detection of dark matter in underground experiments such as XENON100 \cite{xenon100} or LUX \cite{lux}. For instance, the effective interaction between scalar dark matter and gluons is given \cite{gmdm} by
\bea
{\cal L}_{S-N}=\xi_g \, S^2 G_{\mu\nu} G^{\mu\nu}, \quad\quad\xi_g\equiv \frac{c_3 c_S}{ 6\Lambda^2}\,\frac{m^2_S}{m^2_G}.
\eea
Then, the spin-independent cross section induced by the gluon interactions is 
\bea
\sigma_{S-N}= \frac{\mu^2}{\pi m^2_S} \left(\frac{8\pi}{9\alpha_S}\right)^2 m^2_N \xi^2_g f^2_{TG}
\eea
where $\mu=m_S m_N/(m_S+m_N)$ is the reduced mass of the nucleon-dark matter system and 
\bea
f_{TG}= \frac{1}{m_N} \langle N| \frac{-9\alpha_S}{8\pi} G_{\mu\nu} G^{\mu\nu}| N\rangle.
\eea
The lattice result gives $f_{TG}=0.867$ \cite{lattice} while the MILC results ranges between $0.472$ and $0.952$ \cite{MILC}.  For instance, for $m_G=750\,{\rm GeV}$, $m_{\rm DM}=373\,{\rm GeV}$, and $c_3=c_S=1$,  the DM-nucleon scattering cross section becomes $\sigma_{S-N}=1.8-7.5\times 10^{-12}\,{\rm pb}$, which is too small to be constrained by current experiments. The DM-nucleon scattering cross section is given for general DM masses in Fig.~\ref{direct}.

For scalar or fermion dark matter, the annihilation cross section into a pair of massless gauge bosons is velocity suppressed, so there is no constraint from indirect detection in this case \cite{gmdm}. 
On the other hand,  for vector dark matter, the annihilation cross section into a pair of gauge bosons is s-wave, so the model can be constrained by Fermi-LAT \cite{fermilat2,dwarfgalaxy} and HESS \cite{hess2013} gamma-ray searches as well as PAMELA and AMS-02  anti-proton data \cite{antiproton}. In all the cases of dark matter, when dark matter annihilates into a pair of KK gravitons, the cascade decay of each KK graviton into the SM gauge bosons can also lead to gamma-ray or anti-proton signatures \cite{gmdm}.  The detailed discussion on the indirect detection is beyond the scope of our work, but we just remark that anti-proton bounds do not reach the thermal cross section beyond $m_{\rm DM}\approx 200\,{\rm GeV}$\cite{hambye}, for which the correct relic density is obtained in the model.

\section{Kinematic distributions for the spin-two particle}

We have already discussed how the spin-two mediator has a different (more steep) dependence on the energy of the collider, leading to a larger ratio of signal strength from Run1 to Run2. Another interesting aspect of the diphoton signal is the angular distributions, which are very sensitive to the spin of the particle. In this section we discuss the kinematic features of the decay products of a spin-two particle, and compare them with the SM backgrounds as well as a spin-zero hypothesis. 

\begin{figure}
\centering
\includegraphics[scale=0.21]{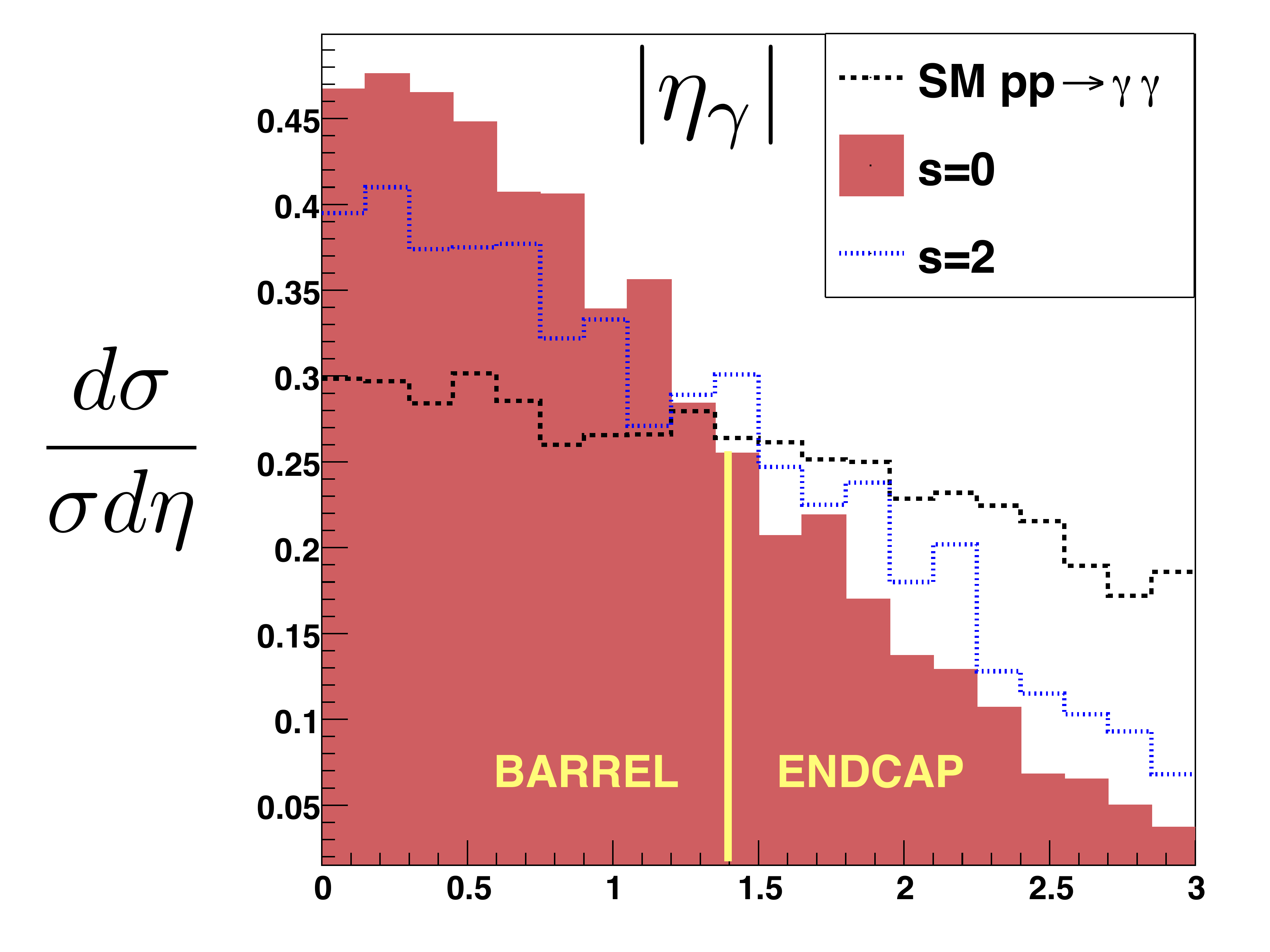}
\includegraphics[scale=0.21]{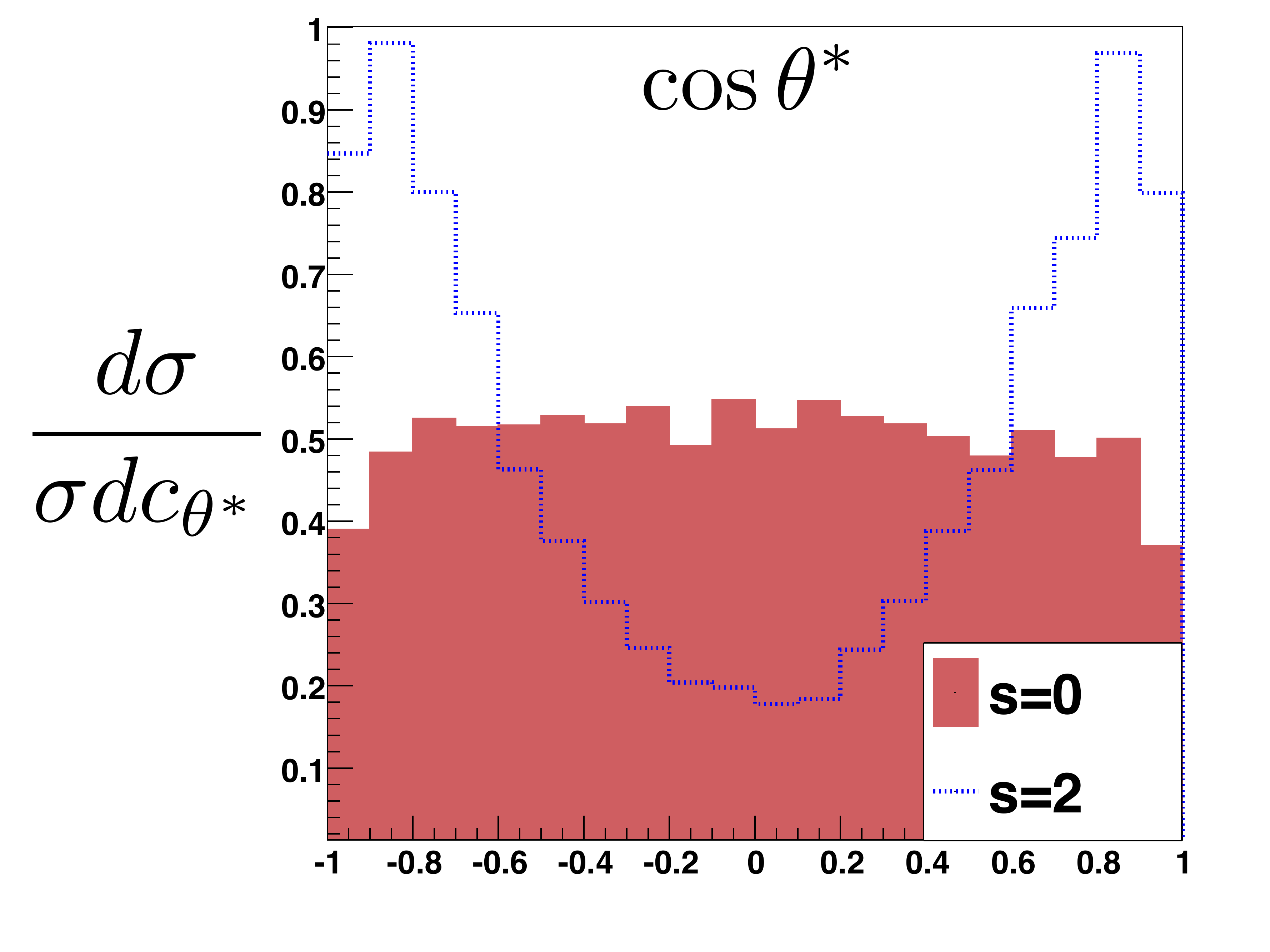}
\caption{{\it  Angular distributions: } (Left) Comparison of the photon rapidity  distribution for the SM, and new heavy spin-zero and -two resonances. (Right) Angle of the photon respect to the beam axis, in the CM frame of the decaying particle. The vertical axes are normalized differential cross sections for comparison. }
\label{fig:distributions}
\end{figure}

A spin-zero particle decaying to two other particles would lead to the symmetric distribution in angle, whereas for a spin-two particle there is a preference to boosted decays. In the left panel of Fig.~\ref{fig:distributions}, we show the rapidity distribution of the photons $\eta_\gamma$ for a scalar and tensor resonance, as well as the SM irreducible background (with a cut on the invariant mass of the diphoton $m_{\gamma\gamma} =750 \pm 50$ GeV) at LHC13. In the right panel, we show the photon angle respect to the beam axis in the CM frame of the decaying particle $\theta^*$, a variable which had been identified in Ref.~\cite{Johngamgam} as the most suitable to gain information on the spin of the Higgs-candidate. A combination of current Run1 and Run2 data could be sufficient to determine the spin of the resonance by performing an analysis on the variable $\cos \theta^*$.

\section{Conclusions}
We have considered the hypothesis that the  diphoton resonance at $750\,{\rm GeV}$ hinted  from both ATLAS and CMS could be a spin-two particle, such as a KK graviton in warped extra dimension, and play the role of a dark matter mediator. This resonance would have direct couplings to all SM particles and Dark Matter. In the extra-dimensional view, the couplings are gravity-induced and dependent on the localization of particles in the extra-dimension. In the dual, strongly-coupled scenario, the couplings would be a function of the degree of compositeness of the particles. We found a large region in the parameter space of the couplings of the KK graviton to the SM gauge bosons capable of explaining the diphoton resonance and at the same time being compatible with the limits from other searches at LHC Run 1. 
Although the ATLAS best-fit slightly prefers a wide width of the diphoton resonance, it is important to combine with the CMS data in favour of a narrow width. We will have to wait for the update of the next-year Run to obtain a more precise measurement estimate of the width.  Since dark matter annihilates into the SM particles through the coupling to the KK graviton, the invisible decay of the KK graviton can be sizeable, depending on the spin of dark matter. Nevertheless, we find that the considered model is not constrained by mono-jet searches in the region of the correct relic density. 

The main purpose of this paper is to present a benchmark for the spin-two hypothesis, which could also provide a natural solution for Dark Matter. Current and future data will allow to determine the spin of the diphoton resonance via angular distributions of the photons. Additionally, one would expect sensitivity to this resonance in other channels, such as dibosons, which can also be used to determine the resonance's quantum numbers.

\section*{Acknowledgments}
We would like to thank Ken Mimasu for conversations on the work. CH appreciates hospitality to IBS-CTPU during his visit. 
The work of CH is supported by the World Premier International Research Center Initiative (WPI Initiative), MEXT, Japan. MP is supported by IBS under the project code, IBS-R018-D1.
The work of HML is supported in part by Basic Science Research Program through the National Research Foundation of Korea (NRF) funded by the Ministry of Education, Science and Technology(2013R1A1A2007919). The work of VS is supported by the Science Technology and Facilities Council (STFC) under grant number ST/J000477/1.




\end{document}